\documentclass{iopart}
\usepackage{graphicx}

\newcommand{\ud}{\mathrm{d}}

\begin{document}
\title{Molecular alignment using circularly polarized laser pulses}
\author{C T L Smeenk and P B Corkum}
\address{JASLab, University of Ottawa and National Research Council, 100 Sussex Drive, Ottawa, Canada}
\ead{christopher.smeenk atsymbol mbi-berlin.de}

\begin{abstract}
  We show that circularly polarized femtosecond laser pulses produce field-free alignment in linear and planar molecules. We study the rotational wavepacket evolution of O$_2$ and benzene created by circularly polarized light. For benzene, we align the molecular plane to the plane of polarization. For O$_2$, we demonstrate that circular polarization yields a net alignment along the laser propagation axis at certain phases of the evolution. Circular polarization gives us the ability to control alignment of linear molecules outside the plane of polarization, providing new capabilities for molecular imaging.
\end{abstract}

Control of molecular alignment and orientation is an essential element of many contemporary experiments in molecular imaging and reaction dynamics \cite{Stapelfeldt2003}. There are two general approaches to alignment -- adiabatic and non-adiabatic. State of the art techniques use supersonic molecular beams and a state selector to direct molecules of a single ro-vibrational state into an ultra-high vacuum chamber \cite{Schnell2009,Filsinger2009a}. By applying a strong non-resonant laser pulse, either adiabatically or non-adiabatically, the most polarizable axis of the molecule can be aligned in the laser field direction of linearly polarized light \cite{Stapelfeldt2003}. Incorporating an elliptically polarized laser pulse in adiabatic alignment allows us to control the major and minor polarizability axes of the molecule  within the polarization plane \cite{Nevo2009}. 

For many applications it is important to study molecules aligned outside the plane of polarization. For example, we could  record molecular frame photo-electron angular distributions (MF-PADs) of k-aligned linear molecules from a new direction with circularly polarized probe pulses \cite{Smeenk2013a}. This kind of experiment is not possible when molecules are aligned in the polarization plane. We show how k-alignment can be achieved. Thereby, we open a third co-ordinate over which we can control the molecular frame-of-reference.

\begin{figure}[t]
  \centering
  \includegraphics[width=6cm]{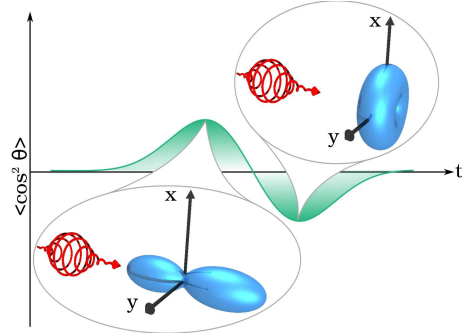}

  \caption{Just prior to the full revival in circularly polarized light (sketched in green), the molecular alignment distribution (inset, blue) is parallel to the propagation axis, $z$. This is ``k-alignment'' (inset left). For a slightly longer pump-probe delay the molecular alignment  is confined to the  plane of polarization. This is ``aplanement'' (inset right). A circularly polarized probe pulse shown in red explodes the molecules to determine the direction of alignment.}
  \label{fig:concept}
\end{figure}

The concept of our experiment is sketched in Fig.~\ref{fig:concept}. We create a rotational wavepacket in O$_2$ using a moderately intense circularly polarized aligning pulse. To determine the dynamics of the wavepacket, including its k-alignment, we use a second high intensity circularly polarized pulse to Coulomb explode the molecule. In a circularly polarized aligning pulse, the most polarizable axis of the molecule receives a torque into the plane of polarization ($xy$ -- see Fig.~\ref{fig:concept}). This creates a rotational wavepacket that re-aligns to the plane of polarization every full revival period $T_R$. For a linear molecule, just before the full revival the internuclear axis first aligns along the direction of laser propagation, $z$ (Fig.~\ref{fig:concept}; inset left). We call this ``k-alignment''. The molecules continue to rotate and a short time later the internuclear axis is aligned with the plane of polarization (Fig.~\ref{fig:concept}; inset right). We call this ``aplanement''. Thus 
circularly polarized aligning pulses allow us to control molecular alignment outside the plane of polarization, along the laser k-axis. This completely optical method applies to any molecule that possesses anisotropic polarizability in a single direction.

We also demonstrate the utility of non-adiabatic alignment with circular polarization by using it to control benzene, a planar molecule. In benzene it is useful to concentrate on the vector normal to the plane containing the carbon and hydrogen nuclei. Aplanement of benzene corresponds to aligning the normal vector along the k-axis (Fig.~\ref{fig:concept} left). Anti-aplanement of benzene corresponds to the normal vector de-localized in the plane of polarization (Fig.~\ref{fig:concept} right).

Experimentally, we use circularly polarized 60 (250) fs pulses produced in a Ti:sapphire laser system operating at 800 nm to align O$_2$ (C$_6$H$_6$). The pump laser intensity was $4\times10^{13}$ W/cm$^2$ ($2\times10^{13}$ W/cm$^2$). The pump pulse is focused by an $f/25$ parabolic mirror onto a pulsed, supersonic gas jet (Even-Lavie valve) operating at 500 Hz. To measure O$_2$ alignment we use neat gas at 8.5 bar backing the valve. The jet temperature is estimated to be 5 K. For benzene alignment we use $10^{-4}$ benzene seeded in He at 14 bar backing pressure.

We use a circularly polarized probe pulse to explode the molecules and measure the fragment momentum vector in a velocity map imaging spectrometer (VMI). The VMI records a differential two-dimensional projection of the fragment momentum distribution, $\ud^2 N/\ud p_x \ud p_z$. The probe is focused onto the sample of aligned molecules with the same parabolic mirror at $f/12.5$. The pump-probe delay is precisely controlled by an automated delay stage. By measuring the momentum distributions from exploded O$^{2+}$ fragments, we measured a degree of alignment of $\langle \cos^2 \theta \rangle=0.7$ using a linearly polarized pump pulse. To determine the molecular alignment angle $\theta$ using a circularly polarized pump pulse, we measure the fragment momentum with respect to the $z$ axis, $\cos \theta = p_z/p$ where $p$ is the maximum fragment momentum. The degree of alignment in the wavepacket is then given by the fragment momentum distribution  projected onto the $z$ axis,
\begin{eqnarray}
  \langle \cos^2 \theta \rangle &=& \int_0^\pi \ud \theta \cos^2 \theta \left(\int \frac{ \ud^2 N}{\ud p_x \ud p_z} \ud p_x \right) \nonumber \\
  &=& \int_{-p}^{+p} \frac{\ud p_z}{\sqrt{p^2 - p_z^2}} \frac{p_z^2}{p^2} \left(\int \frac{ \ud^2 N}{\ud p_x \ud p_z} \ud p_x \right) \ .
\end{eqnarray}
Since the VMI records a \mbox{2D} projection of the momentum distribution, we do not know the full \mbox{3D} momentum of each fragment. We estimate $p$ by measuring a momentum distribution of fragments from Coulomb explosion of randomly aligned molecules. This approach likely underestimates the degree of alignment in the rotational wavepacket.

Measuring the fragment momentum with respect to the $z$ axis inverts the typical revival shape compared to measuring $\theta$ with respect to the polarization axis \cite{Stapelfeldt2003,Dooley03}. In a circularly polarized aligning pulse the molecules are initially pulled into the plane of polarization, $\theta=\pi/2$. Hence aplanement corresponds to a minimum in the $\langle \cos^2 \theta \rangle$ alignment parameter, sketched as the green curve in Fig.~\ref{fig:concept}. Anti-aplanement or k-alignment yields fragments aligned parallel to the $z$ axis: $\theta=\left\{0,\pi \right\}$. Therefore k-alignment corresponds to a maximum in the alignment parameter. This is the opposite of the behaviour one observes when measuring the alignment angle with respect to the polarization axis \cite{Stapelfeldt2003,Dooley03}.


In our experiment the explosion pulse is 50 fs long (FWHM). At this duration detecting alignment using Coulomb explosion by an intense infrared pulse introduces a bias into the measurement. The O$^{2+}$ fragments used to determine the molecule's alignment in the lab frame are produced mainly by enhanced ionization \cite{Seideman1995,Ivanov1996,Constant1996}. Enhanced ionization selects those molecules lying in the plane of polarization. Hence the potion of the rotational wavepacket lying outside the plane of polarization is much less likely to produce the highly charged fragments we use to measure alignment. Using Coulomb explosion by an infrared circularly polarized laser pulse therefore means our probing technique is less sensitive to molecules aligned outside the plane of polarization. We will show that we are still able to make a \emph{relative} measurement of aplanement and k-alignment using Coulomb explosion. However, the bias introduced by enhanced ionization prevents us from quantitatively 
determining the 
degree of alignment. This could be partially improved on by using a few cycle explosion pulse \cite{Legare2006}. 

\begin{figure}[t]
  \centering
  \includegraphics[width=\linewidth]{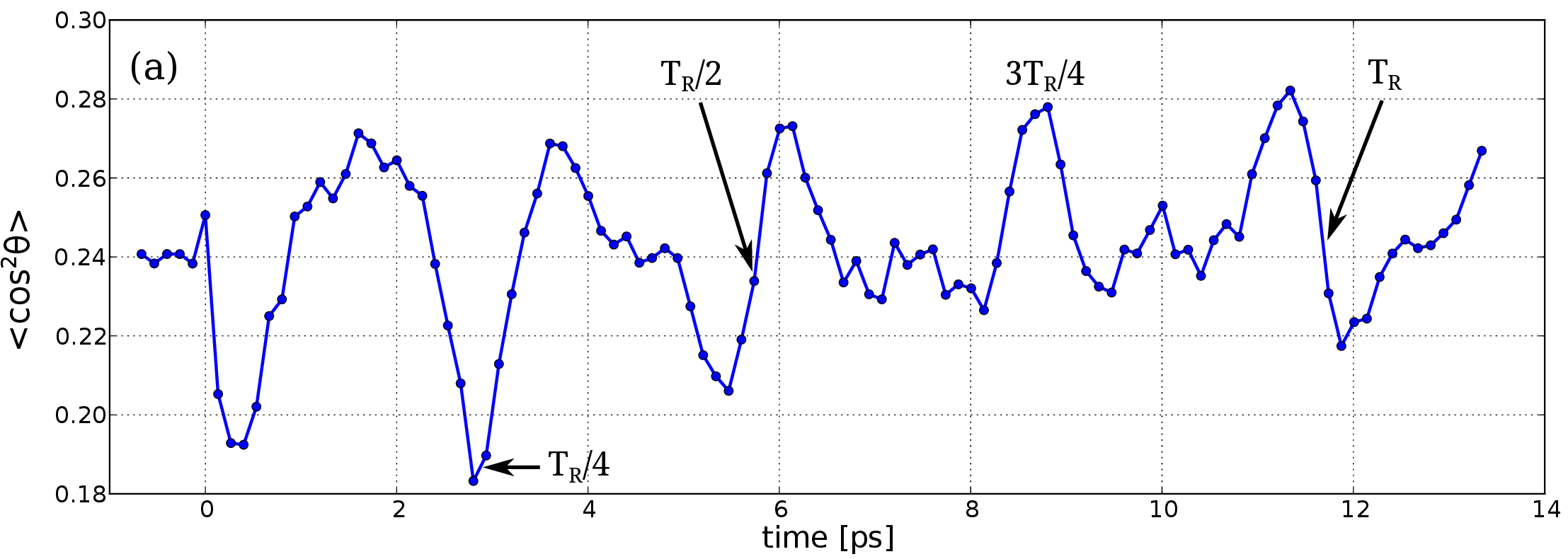} \\
  \includegraphics[width=0.3\linewidth]{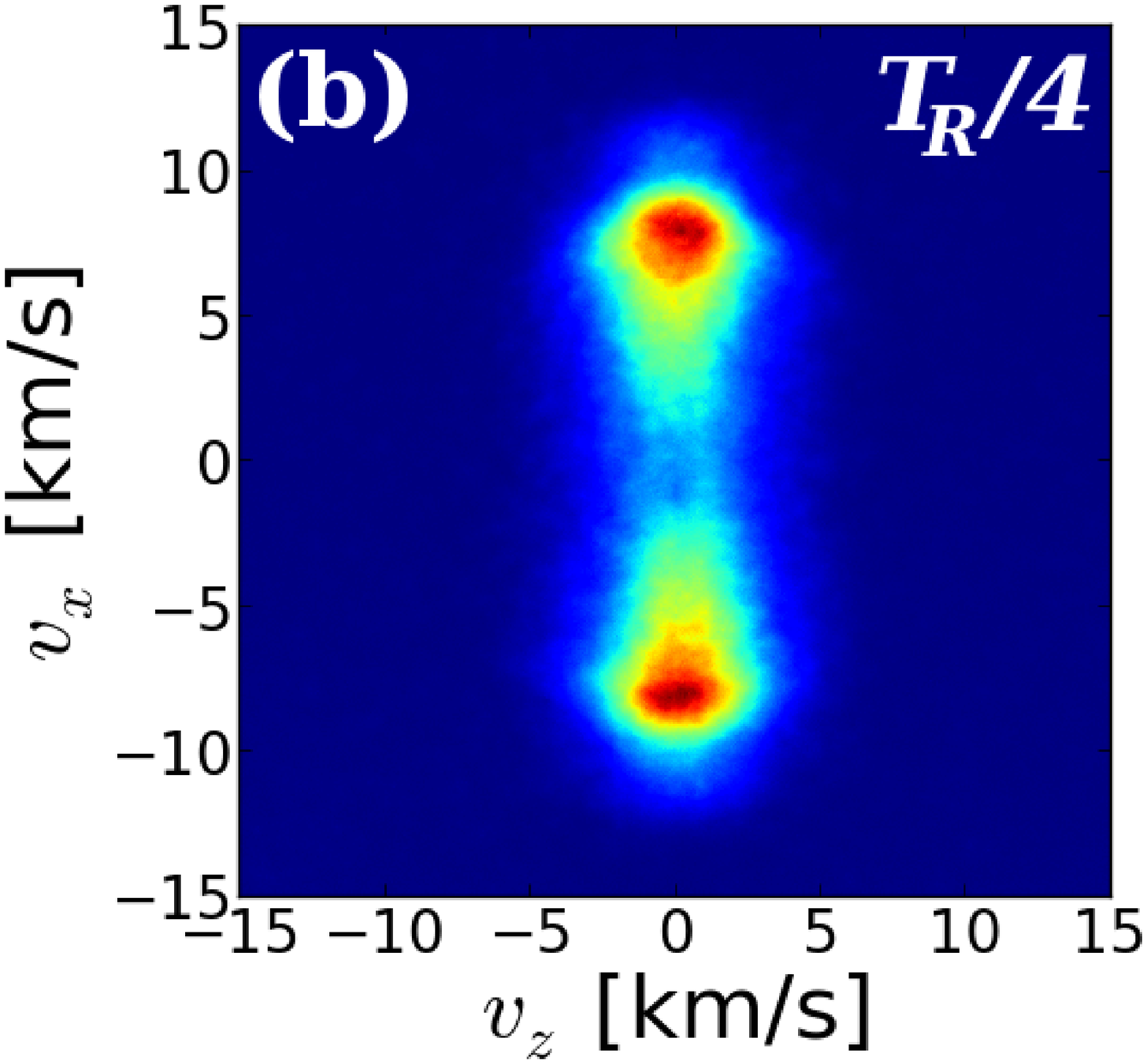}
  \includegraphics[width=0.3\linewidth]{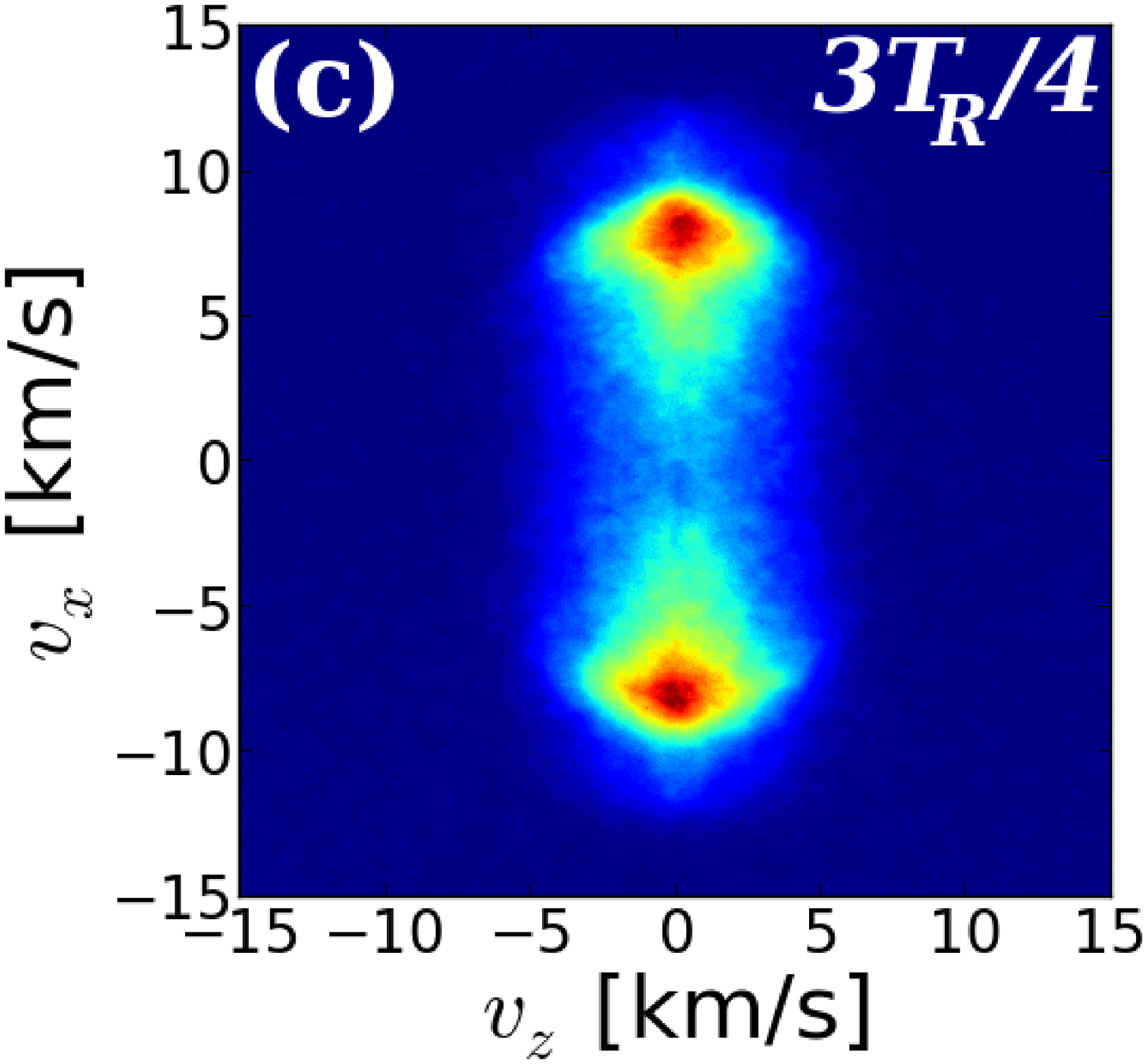}
  \includegraphics[width=0.38\linewidth]{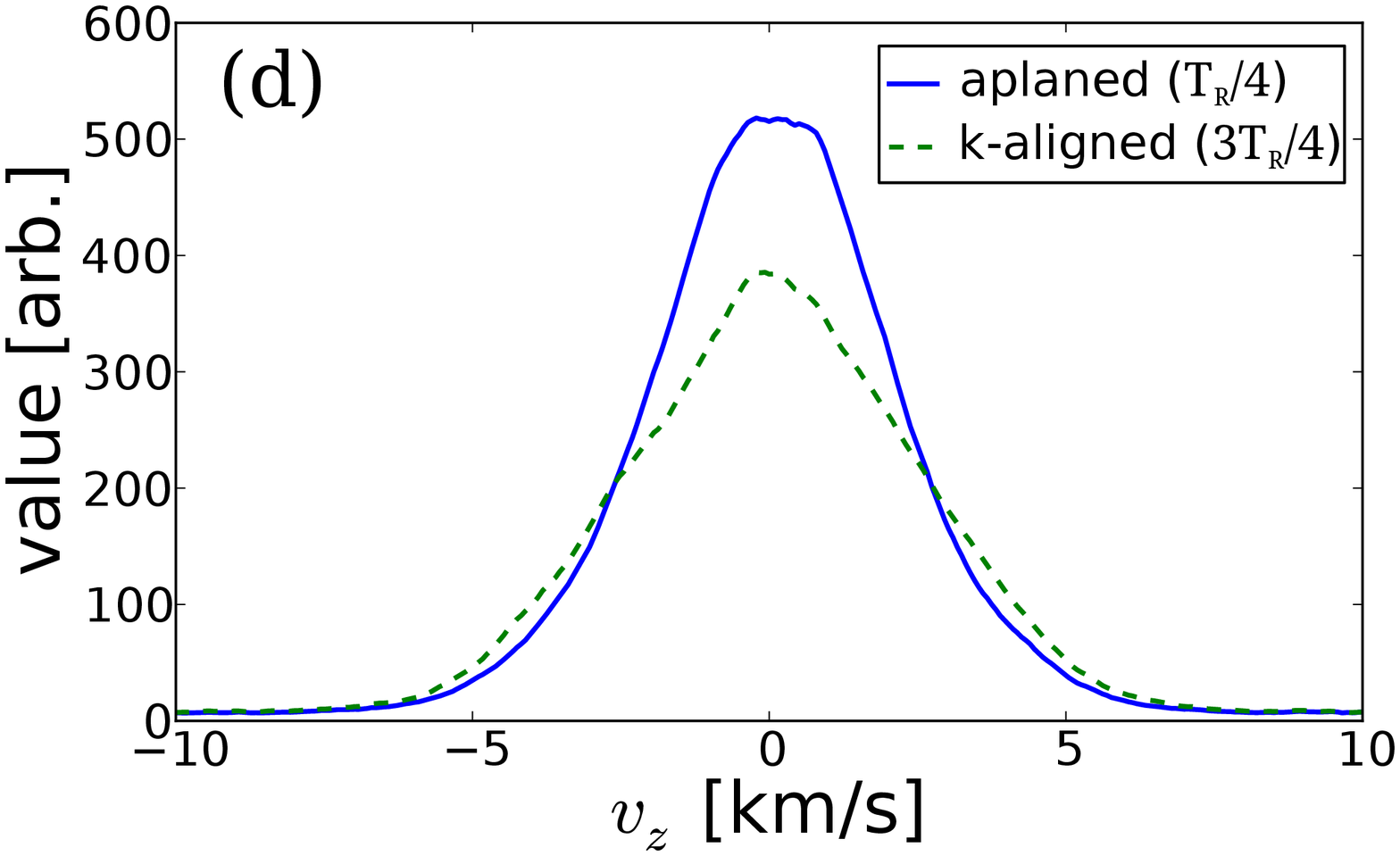}
  \caption{Rotational wavepacket dynamics of O$_2$ using a circularly polarized pump pulse. (a) The wavepacket revival shows strong 1/4, 1/2, 3/4 and full revivals. The full revival period $T_R=11.7$ ps. (b) O$^{2+}$ fragment momentum distribution at aplanement ($T_R/4 =2.90$ ps). (c) O$^{2+}$ fragment momentum distribution at k-alignment ($3T_R/4 =8.68$ ps). (d) Integrated O$^{2+}$ distributions along the propagation axis at aplanement and k-alignment.}
  \label{fig:O2aplanement}
\end{figure}

The rotational wavepacket evolution of O$_2$ created by a circularly polarized pump pulse is shown in Fig.~\ref{fig:O2aplanement}a as seen through the O$^{2+}$ fragments. Initially the molecules are pulled into the plane of polarization, $xy$. Since the angle $\theta$ is measured with respect to the $z$ axis, the alignment parameter $\langle \cos^2 \theta \rangle$ initially decreases relative to the value for unaligned molecules $\approx 0.24$. The decrease in $\langle \cos^2 \theta \rangle$ during the first 250 fs corresponds to aplanement of O$_2$. Subsequently, the wavepacket evolution has a similar temporal behaviour as the rotational wavepacket excited in O$_2$ by a linearly polarized pulse  \cite{Dooley03}. The quarter revival (2.90 ps) exhibits strong aplanement, while at the three-quarter revival (8.68 ps) molecules are preferentially aligned along the k-axis. The strong quarter and three-quarter revivals in O$_2$ are well known and appear because oxygen has nuclear spin $I=0$ and therefore only odd 
rotational states are populated in O$_2$ \cite{Dooley03}. The full revival occurs at $T_R=11.70$ ps just as with a linearly polarized pump pulse. At the full revival anti-aplanement (i.e.~k-alignment) is immediately followed by aplanement. This behaviour is the inverse of the half revival (5.85 ps) where aplanement precedes k-alignment. Each of these temporal features was seen before in alignment using linearly polarized light \cite{Dooley03}. By using a circularly polarized aligning pulse we have observed alignment in a new direction: along the laser k-axis.

The O$^{2+}$ momentum distributions provide direct evidence of aplanement and k-alignment (Figs.~\ref{fig:O2aplanement}b and \ref{fig:O2aplanement}c). The fragment momentum spectra are dominated by enhanced ionization and therefore the maximum signal is always in the plane of polarization. However, the O$^{2+}$ distribution at the 3/4 revival (Fig.~\ref{fig:O2aplanement}c) is extended along the $z$ axis. This is the signature of net k-alignment. Conversely the fragment distribution at 1/4 revival (Fig.~\ref{fig:O2aplanement}b) is more narrowly confined to the plane of polarization. This is due to molecular aplanement at the quarter revival. The contrast between aplanement and k-alignment is more clear when comparing the projection of the respective distributions onto the k-axis (Fig.~\ref{fig:O2aplanement}d). It is clear the distribution at k-alignment is significantly more extended along the $z$ axis.

Our circularly polarized pump pulse produces a torque on linear molecules because of a polarizability asymmetry along one of the principal axes. This pulls the molecular axis into the plane of polarization. A similar polarization asymmetry will occur in planar molecules, for example, benzene. In benzene the polarizability along the two principal axes in the molecular plane is larger than along the axis normal to the molecular plane \cite{Miller1990,Smith2004}. This asymmetry means the dipole induced by a strong non-resonant laser pulse will be larger in the molecular plane than along the normal axis. This leads to a net torque on the molecule and alignment of the molecular plane with the plane of polarization (aplanement). At aplanement the normal vector lies parallel to the k-axis. As the molecule continues to rotate the molecule anti-aplanes. At anti-aplanement the normal vector lies in the plane of polarization.

\begin{figure}
  \centering
  \includegraphics[width=\linewidth]{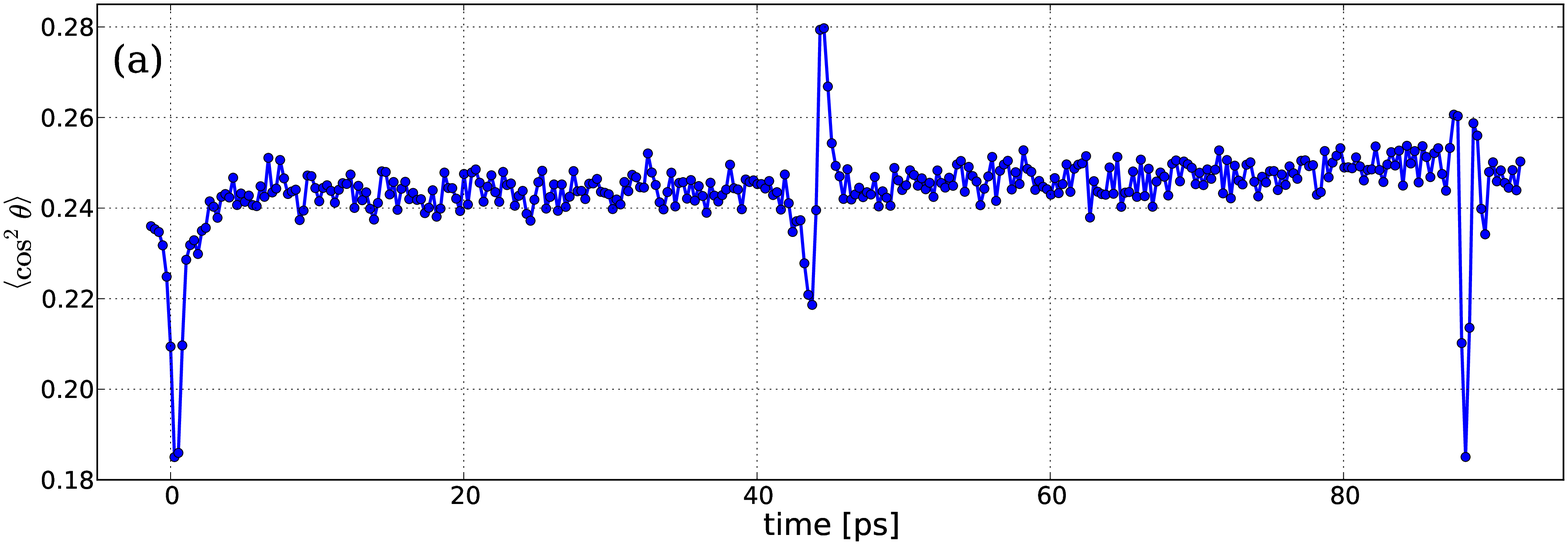} \\
  \includegraphics[width=0.3\linewidth]{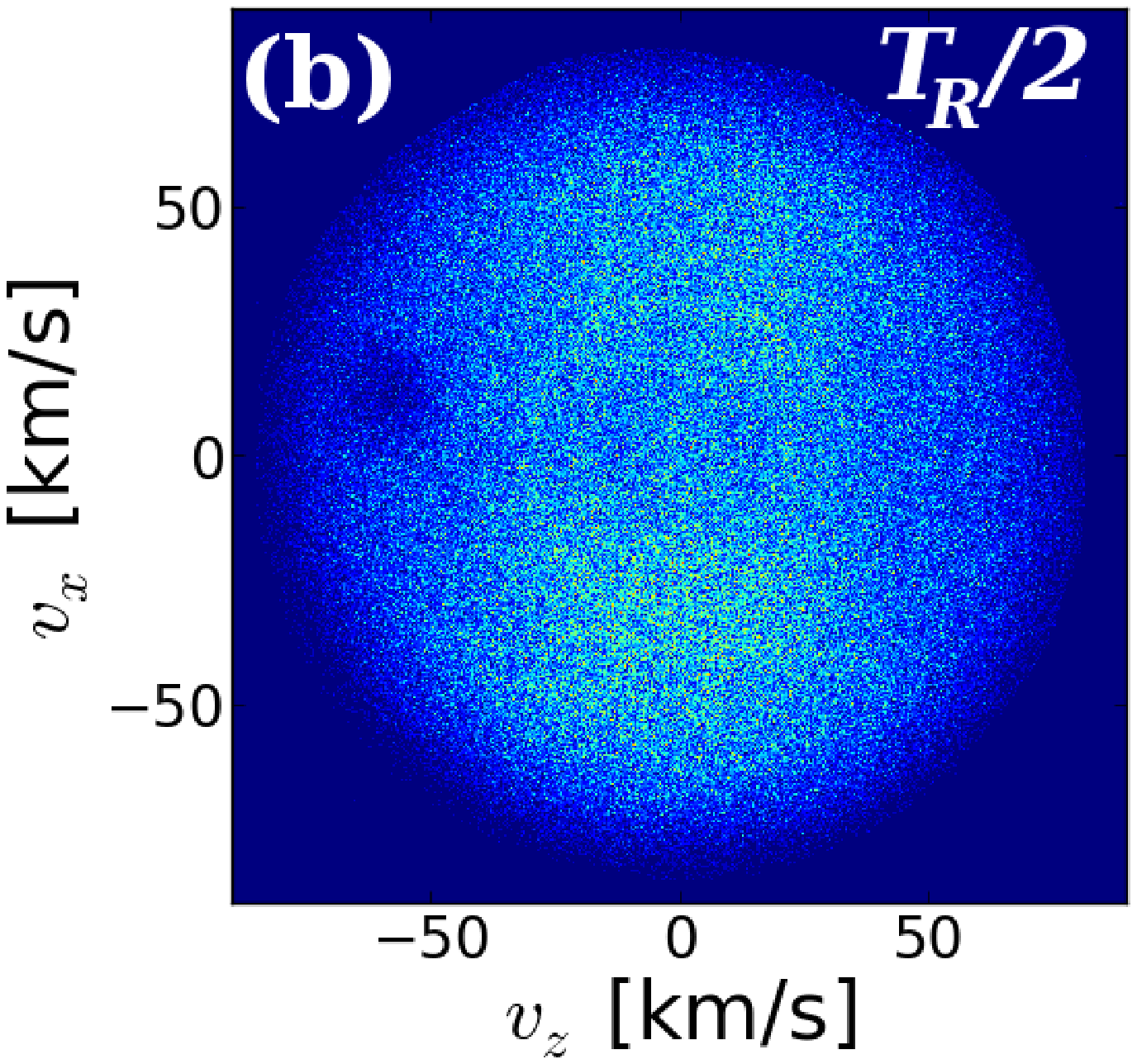}
  \includegraphics[width=0.3\linewidth]{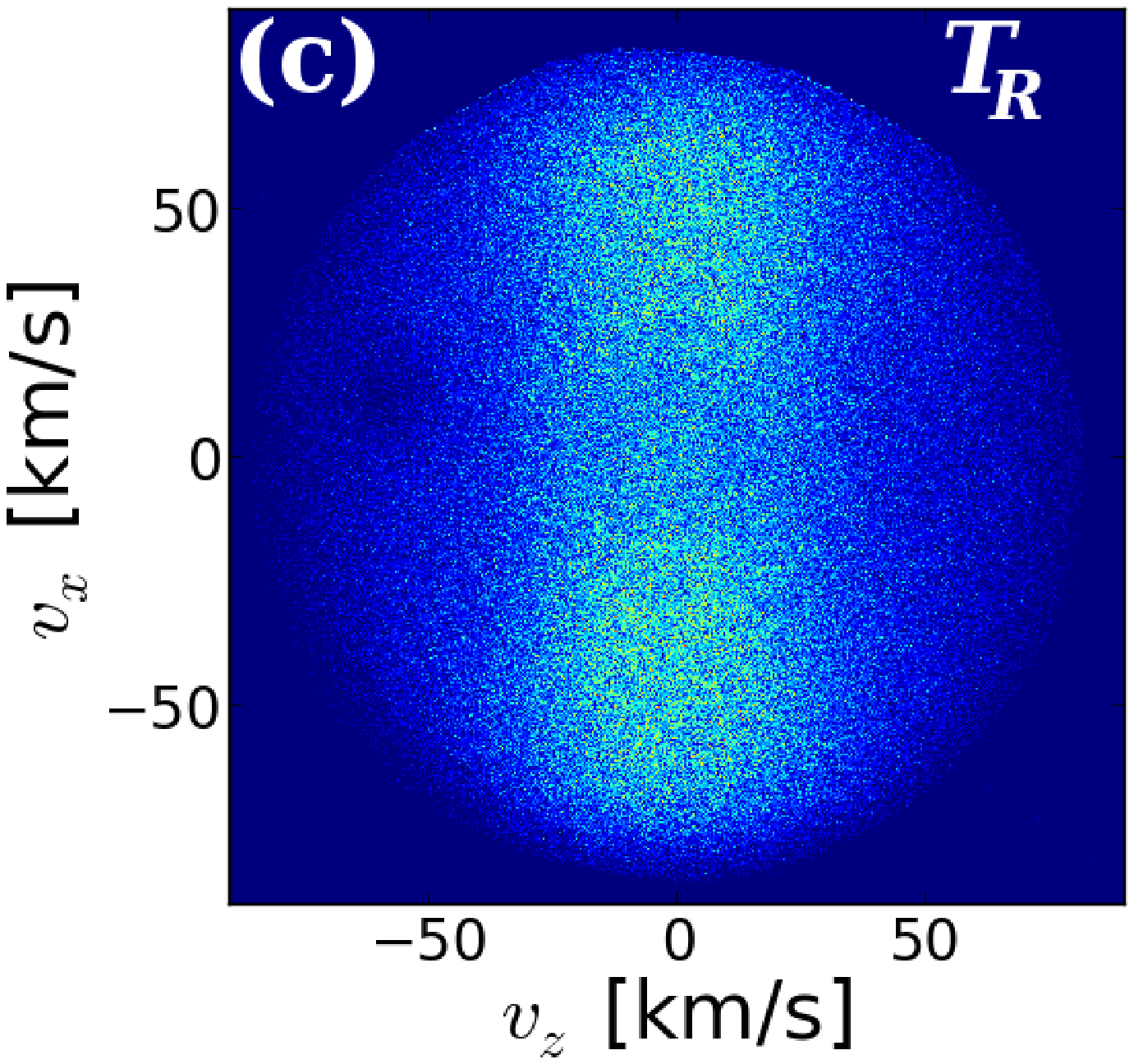}
  \includegraphics[width=0.38\linewidth]{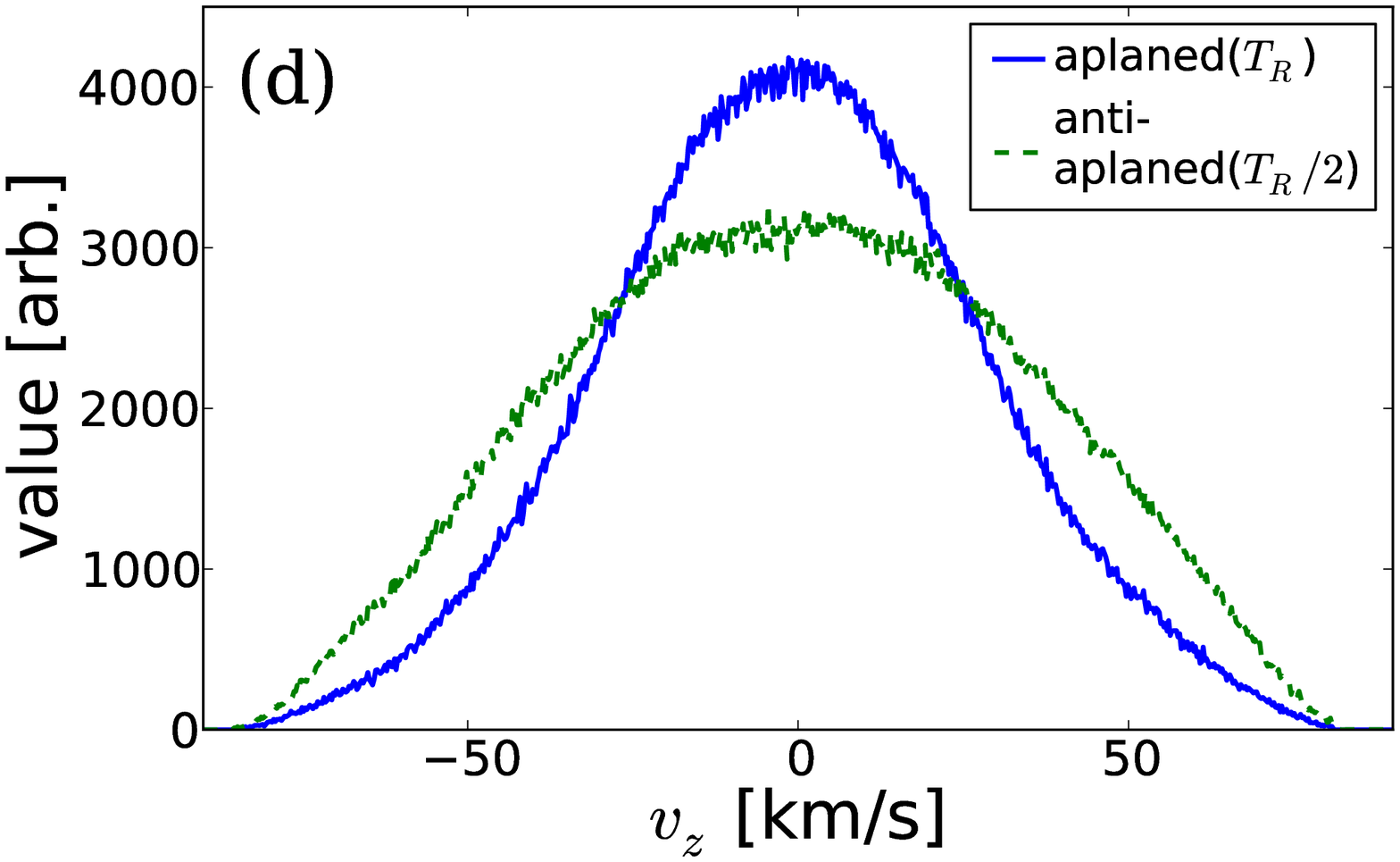}
  \caption{Rotational wavepacket dynamics of C$_6$H$_6$ using a circularly polarized pump pulse. (a) The wavepacket revival shows half revival at 44.1 ps  and full revival at 88.0 ps. (b) H$^+$ momentum distribution at anti-aplanement (44.4 ps). (c) H$^+$ momentum distribution at aplanement (88.4 ps). (d) Integrated H$^+$ distributions along the propagation axis at aplanement and anti-aplanement.}
  \label{fig:benzeneRevival}
\end{figure}

During the Coulomb explosion of benzene, H$^+$ fragments are produced with a broad angular distribution but maximum emission probability in the molecular plane \cite{Shimizu2002,Hansen2011b}. We use the direction of H$^+$ emission as a surrogate for the direction of the molecular plane. By recording the H$^+$ momentum distribution as a function of pump-probe delay, we can watch the rotational wavepacket evolution (Fig.~\ref{fig:benzeneRevival}a). In Fig.~\ref{fig:benzeneRevival}a the circularly polarized pump pulse initially pulls the molecules into the plane of polarization, leading to a decrease in $\langle \cos^2 \theta \rangle$ at prompt alignment. At 44.1 ps the half revival occurs with anti-aplanement maximizing at 44.4 ps. In Fig.~\ref{fig:benzeneRevival}a the full revival occurs at 88.0 ps. We do not observe quarter revivals in benzene. At anti-aplanement, the exploded protons are emitted virtually isotropically from anti-aplaned molecules (Fig.~\ref{fig:benzeneRevival}b) \footnote{In Figs.~\ref{fig:benzeneRevival}b and \ref{fig:benzeneRevival}c there is a small detector imperfection that reduces signal in the upper left hand corner, however, this does not strongly affect the measurement since it is at the edge of the distribution.}. This is in contrast to the proton distribution from aplaned molecules at 88.4 ps where the protons are confined more narrowly to the plane of polarization (Fig.~\ref{fig:benzeneRevival}c).  The contrast between aplanement and anti-aplanement is more clear when viewing the proton distributions projected onto the k-axis (Fig.~\ref{fig:benzeneRevival}d). We observe a substantially narrower proton distribution at 88.4 ps from Coulomb explosion of aplaned benzene. Thus, circularly polarized light allows us to create field-free alignment of a planar molecule.

In conclusion, we have shown how to gain added control over non-adiabatically aligned molecules. Controlling molecular alignment with circular polarization allows us to achieve two goals: (1) we can create field-free alignment of planar molecules in the plane of polarization, (2) we can create a net alignment of linear molecules outside the plane of polarization (i.e.~k-alignment). This allows molecular structure to be studied from a new direction. Although we used Coulomb explosion to observe the alignment, our all-optical technique is not limited to any specific probing method. It can easily be used with higher density gas sources such as those needed for HHG, XUV transient absorption, or ultrafast x-ray diffraction \cite{Itatani2004,Goulielmakis2010,Filsinger2011}. All methods for studying ultrafast structure and dynamics  will benefit from the better control over the molecular frame of reference that we have demonstrated.


\begin{thebibliography}{10}

\bibitem{Stapelfeldt2003}
H~Stapelfeldt and T~Seideman.
\newblock Aligning molecules with strong laser pulses.
\newblock {\em Rev. Mod. Phys.}, 75:543, 2003.

\bibitem{Schnell2009}
M~Schnell and G~Meijer.
\newblock Cold molecules: Preparation, applications, and challenges.
\newblock {\em Angew Chem Int Ed}, 48:6010--6031, 2009.

\bibitem{Filsinger2009a}
F~Filsinger, J~K\"upper, G~Meijer, L~Holmegaard, J~H Nielsen, I~Nevo, J~L
  Hansen, and H~Stapelfeldt.
\newblock Quantum-state selection, alignment, and orientation of large
  molecules using static electric and laser fields.
\newblock {\em J Chem Phys}, 131(6):064309, 2009.

\bibitem{Nevo2009}
I~Nevo, L~Holmegaard, J~H Nielsen, J~L Hansen, H~Stapelfeldt, F~Filsinger,
  G~Meijer, and J~K{\"u}pper.
\newblock Laser-induced 3d alignment and orientation of quantum state-selected
  molecules.
\newblock {\em Phys. Chem. Chem. Phys.}, 11:9912--9918, 2009.

\bibitem{Smeenk2013a}
C~T~L Smeenk, L~Arissian, A~V Sokolov, M~Spanner, K~F Lee, A~Staudte, D~M
  Villeneuve, and P~B Corkum.
\newblock Alignment dependent enhancement of the photo-electron cutoff for
  multi-photon ionization of molecules.
\newblock To be published, 2013.

\bibitem{Dooley03}
P.~W. Dooley, I.~V. Litvinyuk, Kevin~F. Lee, D.~M. Rayner, M.~Spanner, D.~M.
  Villeneuve, and P.~B. Corkum.
\newblock Direct imaging of rotational wave-packet dynamics of diatomic
  molecules.
\newblock {\em Phys. Rev. A}, 68(2):023406, 2003.

\bibitem{Seideman1995}
T~Seideman, M~Y Ivanov, and P~B Corkum.
\newblock Role of electron localization in intense-field molecular ionization.
\newblock {\em Phys. Rev. Lett.}, 75:2819--2822, Oct 1995.

\bibitem{Ivanov1996}
M.~Ivanov, T.~Seideman, P.~Corkum, F.~Ilkov, and P.~Dietrich.
\newblock Explosive ionization of molecules in intense laser fields.
\newblock {\em Phys. Rev. A}, 54:1541--1550, Aug 1996.

\bibitem{Constant1996}
E~Constant, H~Stapelfeldt, and P~B Corkum.
\newblock Observation of enhanced ionization of molecular ions in intense laser
  fields.
\newblock {\em Phys Rev Lett}, 76(22):4140, May 1996.

\bibitem{Legare2006}
F~L{\'e}gar{\'e}, K~F Lee, A~D Bandrauk, D~M Villeneuve, and P~B Corkum.
\newblock Laser coulomb explosion imaging for probing ultra-fast molecular
  dynamics.
\newblock {\em J Phys B}, 39(13):S503, 2006.

\bibitem{Miller1990}
K~J Miller.
\newblock Calculation of the molecular polarizability tensor.
\newblock {\em J Am Chem Soc}, 112:8543, 1990.

\bibitem{Smith2004}
S~M Smith, A~N Markevitch, D~A Romanov, X~Li, R~J Levis, and H~B Schlegel.
\newblock Static and dynamic polarizabilities of conjugated molecules and their
  cations.
\newblock {\em J Phys Chem A}, 108(50):11063--11072, 2004.

\bibitem{Shimizu2002}
S~Shimizu, V~Zhakhovskii, F~Sato, S~Okihara, S~Sakabe, K~Nishihara, Y~Izawa,
  T~Yatsuhashi, and N~Nakashima.
\newblock Coulomb explosion of benzene induced by an intense laser field.
\newblock {\em J Chem Phys}, 117(7):3180--3189, 2002.

\bibitem{Hansen2011b}
J~L Hansen, L~Holmegaard, L~Kalh\o{}j, S~L Kragh, H~Stapelfeldt, F~Filsinger,
  G~Meijer, J~K\"upper, D~Dimitrovski, M~Abu-samha, C~P~J Martiny, and L~B
  Madsen.
\newblock Ionization of one- and three-dimensionally-oriented asymmetric-top
  molecules by intense circularly polarized femtosecond laser pulses.
\newblock {\em Phys. Rev. A}, 83:023406, Feb 2011.

\bibitem{Itatani2004}
J.~Itatani, J.~Levesque, D.~Zeidler, H.~Niikura, H.~Pepin, J.~C. Kieffer, P.~B.
  Corkum, and D.~M. Villeneuve.
\newblock Tomographic imaging of molecular orbitals.
\newblock {\em Nature}, 432(7019):867--871, 2004.

\bibitem{Goulielmakis2010}
E~Goulielmakis, Z~H Loh, A~Wirth, R~Santra, N~Rohringer, V~S Yakovlev,
  S~Zherebtsov, T~Pfeifer, A~M Azzeer, M~F Kling, S~R Leone, and F~Krausz.
\newblock Real-time observation of valence electron motion.
\newblock {\em Nature}, 466(7307):739, 2010.

\bibitem{Filsinger2011}
F~Filsinger, G~Meijer, H~Stapelfeldt, H~N Chapman, and J~K{\"u}pper.
\newblock State- and conformer-selected beams of aligned and oriented molecules
  for ultrafast diffraction studies.
\newblock {\em Phys. Chem. Chem. Phys.}, 13:2076--2087, 2011.

\end{thebibliography}

\end{document}